\documentclass[aps,pra,reprint, amsmath, amssymb,superscriptaddress]{revtex4-1}
 
\usepackage{bm}
\usepackage[retainorgcmds]{IEEEtrantools}
\usepackage{graphicx}
\usepackage{mathrsfs}
\usepackage{amsmath}
\usepackage{amsfonts}
\usepackage{amssymb}
\usepackage{color}
\usepackage{times,txfonts}
\usepackage{nicefrac}
\usepackage{ragged2e}
\usepackage{tikz}

\usepackage[colorlinks=true,linkcolor=blue,urlcolor=blue,citecolor=blue,pdfusetitle]{hyperref}

\usepackage{blindtext}

\newcommand{\tr}{\mathrm{tr}}

\definecolor{cadmiumgreen}{HTML}{097969}

\begin{document}

\title{Battery charging in collision models with Bayesian risk strategies}
\date{\today}
\author{Gabriel T. Landi}
\email{gtlandi@gmail.com}
\affiliation{Instituto de F\'isica da Universidade de S\~ao Paulo,  05314-970 S\~ao Paulo, Brazil.}
\affiliation{School of Physics, Trinity College Dublin, College Green, Dublin 2, Ireland}

\begin{abstract}
We construct a collision model where measurements in the system, together with a Bayesian decision rule, are used to classify the incoming ancillas as having either  high or low ergotropy  (maximum extractable work). 
The former are allowed to leave, while the latter are redirected for further processing, aimed at increasing their ergotropy further. 
The ancillas play the role of a quantum battery, and the collision model therefore implements a Maxwell demon. 
To make the process autonomous, and with a well defined limit cycle, the information collected by the demon is reset after each collision by means of a cold heat bath. 
\end{abstract}

\maketitle{}

\section{Introduction}

Collision models, first studied in the seminal paper by Rau~\cite{Rau1963}, have seen a revival of interest in recent years~\cite{Ziman2002,Scarani2002,Englert2002}. 
They replace the complex system-bath dynamics by a series of sequential collisions between a system of interest and a continuous stream of small units, called ancillas. 
This not only makes the dynamics simpler, but also more controllable. 
For example, collisional models have proven to be crucial in developing the basic laws of thermodynamics in the quantum regime~\cite{DeChiara2018,Landi2020a,Barra2015,Strasberg2016}, or to further our understanding of non-Markovianity~\cite{Campbell2019a,Man2018,Lorenzo2017,Donvil2021,Mascarenhas2017,McCloskey2014,Campbell2018b,Rybar2012a,Cilluffo,Ciccarello2013b,Bernardes2017,Taranto2018a,Ciccarello2013a,Filippov2017,Kretschmer2016,Jin2018,Ciccarello2013,Bernardes2014,Cakmak2017,Daryanoosh2018,DeChiara2020}. 
For a recent review, see~\cite{Ciccarello2020}.

A particularly nice feature of these models is that they allow for a clean implementation of autonomous processes: 
Ancillas arrive, undergo some physical process, and then leave. 
Different implementations can  be used to perform different tasks, which are gauged by the changes in the ancilla's state.
Moreover, the process is allowed to continue indefinitely, as long as new ancillas continue to arrive.
Indeed, there have already been several proposals which employ collision models in e.g. quantum heat engines~\cite{Quan2007,Allahverdyan2010,Uzdin2014,Campisi2014,Campisi2015,Denzler2019,Pezzutto2019,Mohammady2019,Molitor2020} or quantum thermometers~\cite{Seah2019,Shu2020,Alves2021}.

In this paper we discuss the implementation of an autonomous collision model engine aimed at charging quantum batteries.
Battery charging in the quantum domain is an active field of study~\cite{Skrzypczyk2013,Campaioli2017,Teo2017,Andolina2019,Mitchison2021}. 
The present framework aims to produce a model in which this charging occurs autonomously, for an arbitrary number of charging units, and in a way which works for arbitrary initial battery states.

\begin{figure*}
    \centering
    \includegraphics[width=\textwidth]{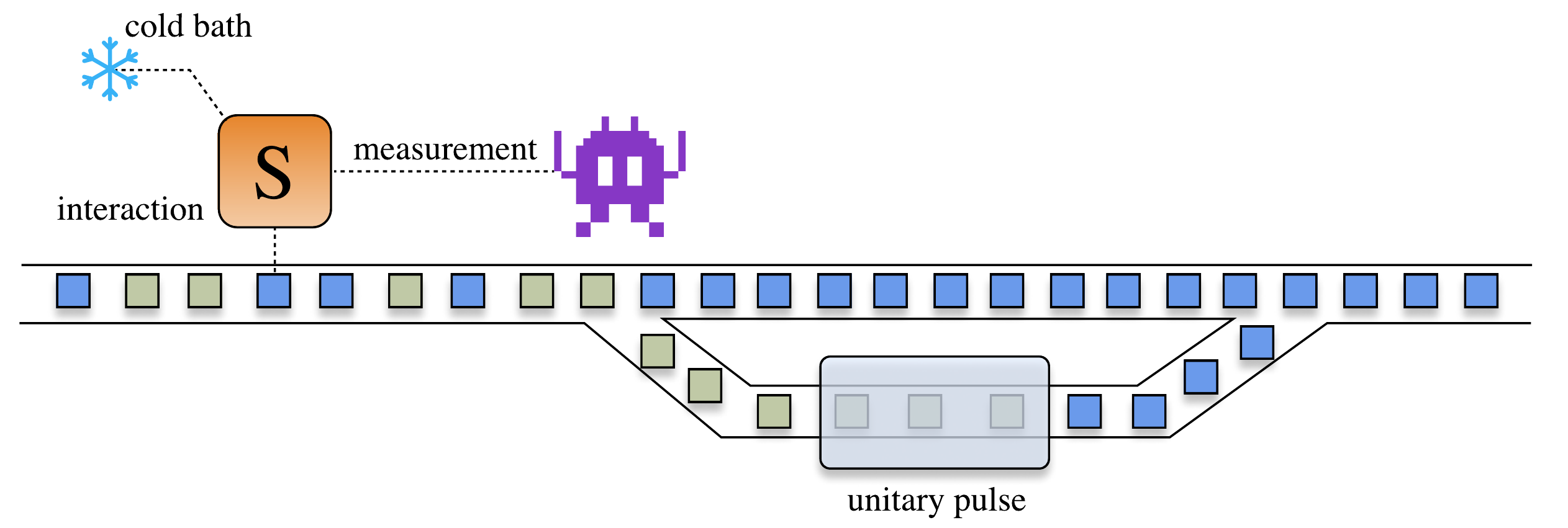}
    \caption{Autonomous collision model for enhancing the ergotropy in an ensemble of ancillas. 
    A stream of ancillas, drawn from random states, interact with a system $S$. Measurements in $S$ are then used to distinguish whether the ancillas have low or high ergotropy. 
    This information is used by a (space invader) demon, operating under the paradigm of Bayesian Decision Theory, to decide whether or not the ancillas should be further processed or not, with the goal of increasing their ergotropy even further.
    }
    \label{fig:drawing}
\end{figure*}

The input of the engine is a stream of ancillas, drawn randomly from some ensemble of states.
The thermodynamic ``usefulness'' of each ancilla will be characterized by its ergotropy~\cite{Allahverdyan2004}, which quantifies the maximum amount of work that can be extracted from it by means of a unitary interaction. 
The goal of the engine is then to increase the average ergotropy of the outgoing ancillas.
This is accomplished by using information extracted from measurements in the system, as depicted in Fig.~\ref{fig:drawing} (the ancillas are never measured). 
This setup was inspired by Ref.~\cite{Francica2016a}, which studied the ergotropy that could be extracted from quantum correlations between a system and a single ancilla. 
And it is  opposite in spirit to, e.g., continuously monitored systems~\cite{Wiseman2009,Jacobs2014}, where one uses measurements in the ancillas to learn something about the system~\cite{Gross2017a,Rossi2020,Landi2021a,Belenchia2019}; here we use instead information about the system to learn about the ancillas. 

The measurement outcomes are used to classify the ancillas as having either high or low ergotropy, which we model  using Bayesian decision theory~\cite{Duda2000}.
This therefore implements a Maxwell demon~\cite{Maxwell1888}, which autonomously decides what to do with each ancilla. 
High ergotropy ancillas (defined according to some threshold) are allowed to leave, while low ergotropy ones are flagged for further processing. 
That is, they are redirected to go through another quantum channel, aimed at increasing their ergotropy further (Fig.~\ref{fig:drawing}). 
In our case, we will model this in terms of an additional unitary pulse, but more general quantum channels can also be used. 

The system in this case plays the role of a memory. 
As is well known, the process of acquiring information can in principle be done without any energetic cost. 
However, there is a fundamental cost in erasing the information~\cite{Bennett1973,Plenio2001}, given by Landauer's principle~\cite{Landauer1961}.
We model this by assuming that the system is coupled to a cold heat bath that acts for a finite time, in between collisions. 
As we show, this is crucial for the engine to operate autonomously.

\section{Basic model}

We consider a stream of ancillas, each prepared in a state $|\psi_A\rangle$ drawn from an ensemble of $d$ possible states $\{|\psi_i\rangle\}$ (not necessarily orthogonal), with probability $q_i$.
Often, in the collision model literature, one assumes that the ancillas are in mixed states. 
This is a natural choice if one is interested in the steady-state properties of the system. 
But here, for the task at hand, it is much more natural to assume that the ancillas are in pure states. 
Notwithstanding, all results below also hold for ensembles of mixed states.
The notations $\psi_A = |\psi_A\rangle \langle \psi_A|$ will be used whenever the ancilla state is pure. 

The thermodynamic utility of each ancilla can be quantified by its ergotropy~\cite{Allahverdyan2004} which, for a generic ancilla state $\rho_A$, is defined as
\begin{equation}\label{ergotropy}
    \mathcal{W}(\rho_A) = \tr(\rho_A H_A) - \min\limits_{V} \tr(\rho_A V^\dagger H_A V),
\end{equation}
where $H_A$ is the ancilla Hamiltonian and the minimization is over all unitaries $V$. When the state is pure, this reduces to the more intuitive result 
\begin{equation}\label{ergotropy2}
    \mathcal{W}(\psi_A) = \langle \psi_A | H_A | \psi_A\rangle - E_{\rm gs}^A,
\end{equation}
where $E_{gs}^A$ is the ground state of $H_A$.

The stream of ancillas are first put to interact with a system $S$, one at a time, for a fixed duration $\tau_{SA}$, according to some Hamiltonian $H_{SA}$.
If the system is in $\rho_S$ and the ancilla is in $\psi_A$, this produces the map
\begin{equation}
    \rho_{SA|\psi_A} = e^{-i H_{SA} \tau_{SA}} (\rho_S \otimes \psi_A) e^{i H_{SA} \tau_{SA}}.
\end{equation}
Immediately afterwards, the system is measured, which we  describe by a set of Kraus operators $\{M_x\}$, with $m$ possible outcomes, $x = 1, \ldots, m$, and satisfying $\sum_x M_x^\dagger M_x = 1$.
The probability of outcome $x$, conditioned on the initial ancilla state, is 
\begin{equation}\label{likelihood}
    P(x|\psi_A) = \tr\Big\{ (M_x^\dagger M_x \otimes I_A )\rho_{SA|\psi_A}\Big\},
\end{equation}
where $I_A$ is the identity acting on the ancilla.
Moreover, if outcome $x$ is observed, the reduced state of $SA$ should be updated to 
\begin{equation}\label{conditional_joint_state}
    \rho_{SA|x,\psi_A} = (M_x \otimes I_A) \rho_{SA|\psi_A} (M_x^\dagger \otimes I_A). 
\end{equation}
From this, the reduced states of system and ancilla, $\rho_{S|x,\psi_A}$ and $\rho_{A|x,\psi_A}$, can be obtained by taking the partial trace.

In between collisions, the state of the system is allowed to relax in contact with a heat bath, which we describe by a Lindblad master equation acting for a fixed time $\tau_{SE}$. 
It is assumed for simplicity that $\tau_{SA} \ll \tau_{SE}$ so that, during the system-ancilla interaction, the system is approximately uncoupled from the bath.

Based on the outcome $x$, a demon tries to classify whether an ancilla has a high or a low ergotropy $\mathcal{W}$ (according to some model-specific threshold).
The former can leave the process, while the latter are redirected for additional processing, aimed at increasing their ergotropy further. 
We describe this in terms of a unitary pulse $\mathcal{O}$, so that the final state of the ancilla will be 
\begin{equation}\label{cases}
    \rho_A' = \begin{cases}
    \rho_{A|x,\psi_A} & \text{high ergotropy in } \psi_A,
    \\[0.2cm]
    \mathcal{O} \rho_{A|x,\psi_A} \mathcal{O}^\dagger, & \text{low ergotropy in } \psi_A.
    \end{cases}
\end{equation}
The meaning of low or high ergotropy is model specific, and will be discussed further below.
The ultimate goal of the engine is thus to produce an ensemble with average ergotropy higher than that of the initial ensemble $\{q_i, |\psi_i\rangle\}$:
\begin{equation}
    \mathcal{W}_{\rm raw} = \sum\limits_i q_i \mathcal{W}(\psi_i),
\end{equation}
where the subscript ``raw'' will always refer to the ancillas before entering the engine.

\section{Bayesian risk analysis}

Before discussing an actual implementation, we must first discuss the type of rationale that will be used by the demon in deciding whether the ergotropy is high or low. 
We do this using the concept of  Bayesian risk analysis, as a general tool for implementing the decision process. 

There are $d$ possible preparations $\psi_i$, and $m$ possible outcomes $x$, each pair associated to a certain quantum state $\rho_{A|x,\psi_i}$ [Eq.~\eqref{conditional_joint_state}].
It is assumed that the demon knows the possible set of states $\{\psi_i\}$, but does not know the current ancilla state, nor the probabilities $q_i$ with which they were sampled (the latter restriction could be lifted without significantly altering the problem). 
At each collision, all the demon knows is therefore the outcome $x$.
Based on this, it may take one of a set of $a$ actions $\alpha_k$, $k = 1,\ldots, a$. 
Generally speaking, we could associate each action with a quantum channel $\mathcal{E}_k$, which will process the quantum state of the ancilla further. 
For example, in the case of Eq.~\eqref{cases}, action $\alpha_1$ stands for ``do nothing,'' while $\alpha_2$ stands for the unitary channel $\mathcal{O}\bullet \mathcal{O}$. 
But, more generally, all kinds of channels can in principle be used. 

In Bayesian risk analysis, we quantify each action by a certain gain, described by a  non-negative function $\lambda(\alpha_k |x, \psi_i)$ determining how much is gained from using action $\alpha_k$ when the outcome is $x$ and the state is $\psi_i$ (one could equivalently frame the problem in terms of risks, instead of gains). 
This set of function determines the type of strategy the demon will use, and different functions will lead to different engine performances. 
An example could be the
ergotropy~\eqref{ergotropy} of 
$\mathcal{E}_k(\rho_{A_x,\psi_i})$; that is, 
\begin{equation}
    \lambda(\alpha_k |x, \psi_i) = \mathcal{W}\Big(\mathcal{E}_k(\rho_{A_x,\psi_i})\Big)
\end{equation}
However, as we will show below, in specific models simpler functions often be employed.

For each outcome $x$, the demon's decision will then be to choose the action $\alpha_k$ which maximizes the Bayesian gain
\begin{equation}\label{gain}
    G(\alpha_k | x) = \sum\limits_i \lambda(\alpha_k | x, \psi_i) P(\psi_i|x),
\end{equation}
where $P(\psi_i | x)$ is the probability that the initial state was $|\psi_i\rangle$ given that the outcome in the system was $x$. 
According to Bayes's rule, this is further given by 
\begin{equation}\label{bayes_theorem}
    P(\psi_i|x) = \frac{P(x|\psi_i) P(\psi_i)}{\sum\limits_i P(x|\psi_i) P(\psi_i)},
\end{equation}
where $P(x|\psi_i)$ is the likelihood function, given in Eq.~\eqref{likelihood}, and 
$P(\psi_i)$ is the prior probability the demon associates to the ancilla being in $|\psi_i\rangle$.

If the demon does not know in advance how the ancillas are  sampled, the priors $P(\psi_i)$ will in general differ from the $q_i$. In fact, in the beginning of the process, a natural choice of prior would be $P(\psi_i) = 1/d$.
After each collision, however, the demon updates $P(\psi_i)$ to the posterior $P(\psi_i|x)$, which can then be used as the prior for the next step.
Under mild conditions, it is  expected that in the steady state this should converge to the true sampling probabilities $q_i$. 

We also mention that, in general, the state of the system is constantly changing. 
As a consequence, when the above procedure is used sequentially, it may cause $P(\psi_i|x)$ at the $n$-th step to depend on the outcomes of all past collisions, thus making the process highly non-Markovian. 
In fact, even in the limiting case of projective measurements,  $P(\psi_i|x)$ would still depend on the previous outcome.
This is directly associated with Benett's exorcism of Maxwell's demon~\cite{Bennett1973}: while there is no minimum cost to acquire information, there is always a fundamental heat cost for erasing it (see also~\cite{Plenio2001}). 
If the engine is to operate autonomously, the memory (which is in this case the system) must be reset at each step. 
In practice, the demon may continue to employ the same gain function~\eqref{gain}, which would happen when it is unaware of whether the system has been fully reset or not.
The only problem is that this may cause it to make wrong decisions. 
The better is the memory reset, the more accurate is the demon's decision.

\section{Qubit-qubit model}

We now consider a concrete implementation of this approach, where we assume that the system and ancillas are all made of qubits. 
The ancilla Hamiltonian is taken to be $H_A = -\omega \sigma_z^A/2$, where $\sigma_z$ is a Pauli matrix. 
The ground-state is thus the computational basis state $|0\rangle$; i.e., $\sigma_z|0\rangle = |0\rangle$. 
The ergotropy~\eqref{ergotropy} is then bounded between $\mathcal{W} \in [0,\omega]$, with the maximum being for the excited state $|1\rangle$.

The system-ancilla interaction is taken as
\begin{equation}\label{example_HSA}
    H_{SA} = g \sigma_y^S \otimes \sigma_z^A. 
\end{equation}
This is a typical pointer-basis type of measurement~\cite{Zurek1981}, with information on the ancilla's population being directly encoded in the system, while at the same time causing the coherence's to dephase. 
The ergotropy~\eqref{ergotropy} has  contributions from both the populations and  coherences~\cite{Francica2020}. 
The interaction with the system will keep the former intact, but disturb the latter (measurement backaction). 
The goal, therefore, is to see if one can increase the ergotropy from the populations while, at the same time, not excessively harm that from the coherences.

The system is  measured after each step in the eigenbasis $|\pm\rangle = (|0\rangle \pm |1\rangle)/\sqrt{2}$ of the $\sigma_x$ operator. 
To understand why this is a good measurement strategy, suppose that the system is initially prepared in $\rho_S = |0\rangle\langle 0|$, while the ancilla is in  $|\psi_A \rangle = \cos(\theta/2)|0\rangle + e^{i \phi} \sin(\theta/2)|1\rangle$. 
Then Eq.~\eqref{likelihood} will produce the likelihoods
\begin{equation}\label{example_likelihood}
    P(x| \psi_A) = \frac{1}{2} \Big[1 +x \sin(2g \tau_{SA}) \cos\theta\Big].
\end{equation}
For $\theta \in [0,\pi/2]$ (northern hemisphere in Bloch's sphere), the outcome $x = +1$ is more likely, while for $\theta \in [\pi/2,\pi]$ (southern hemisphere) it is actually $x=-1$. 
But the ergotropy is directly related to the position in Bloch's sphere, being low in the former and high in the latter.
This means that if $x=+1$ is observed, it is more likely that the ancilla has a low ergotropy. 
A very simple Bayesian strategy is thus to take the gain of no action ($\alpha_1$) as $\lambda(\alpha_1|x, \psi_i) = 1$ when $x=-1$, and zero otherwise; and similarly  $\lambda(\alpha_2|x, \psi_i) = 1$ when $x = 1$, and zero otherwise. 

When the ancilla is flagged, it is more likely to be in the northern hemisphere. In this case, we can then apply an additional unitary pulse  $\mathcal{O} = \sigma_x^A$, which flips the ancilla's state to the southern hemisphere. 
Note that if the ergotropy is already high, this will generally spoil it. 
That is to say, whenever the demon makes a mistake, it will actually be degrading the ancilla's ergotropy. 
But since correct decisions are more likely, it will on average increase it.

Finally, between measurements the system is taken to interact with a zero temperature heat bath for a time $\tau_{SE}$, described by the master equation 
\begin{equation}\label{QME}
    \frac{d\rho_S}{dt} = -i[H_S, \rho_S] + \gamma  D[\sigma_+^S]\rho_S,
\end{equation}
where $\gamma$ is the coupling strength and $D[L]\rho = L \rho L^\dagger - \frac{1}{2}\{L^\dagger L, \rho\}$. 
Moreover, we assume $H_S = -\omega_S\sigma_z^S/2$, with $\omega_S$ not necessarily resonant with the ancilla frequency $\omega$.

\begin{figure*}
    \centering
    \includegraphics[width=\textwidth]{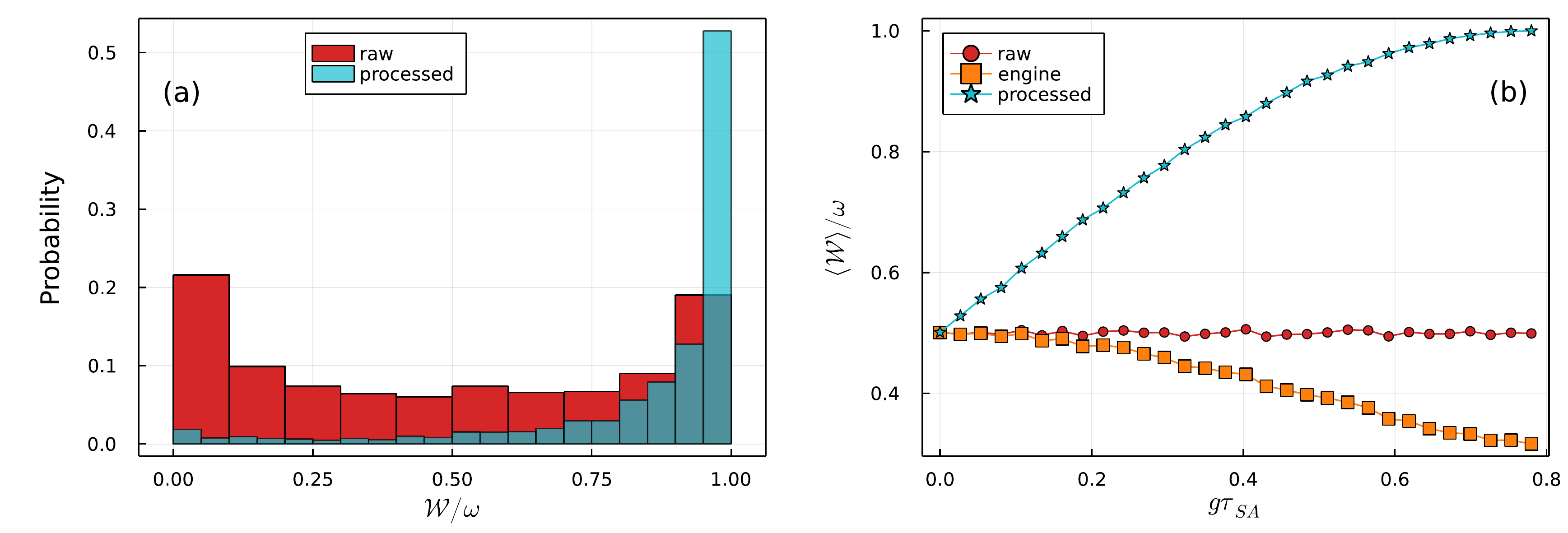}
    \caption{Battery charging in a qubit-qubit collisional model. 
    (a) Histogram of the ergotropies obtained from randomly sampled ancilla states (red), as compared with the final ergotropies after they are passed through the engine. The data was sampled from   $N = 10^4$ simulations, with the system-ancilla interaction strength  fixed at $g\tau_{SA} = \pi/8$.
    (b) Average ergotropy as a function of $g\tau_{SA}$. Raw values (which are independent of $g\tau_{SA}$) and processed values are shown in the same color code as in (a). The points marked as ``engine'' refer to the ergotropy when all ancillas are passed through the engine, irrespective of the outcome $x$.
    }
    \label{fig:example}
\end{figure*}

\section{Results}

In what follows, the ancillas are all uniformly sampled from generic states $|\psi_i\rangle$ within the Bloch sphere, using the appropriate Haar measure.
We start by assuming that $\gamma\tau_{SE}$ is sufficiently large so that, after each step, the state of the system is fully reset back to $\rho_S = |0\rangle\langle 0|$.
Illustrative results are shown in Fig.~\ref{fig:example}. 
The histogram in Fig.~\ref{fig:example}(a) compares the raw ergotropy with that obtained at the output of the engine, for fixed $g\tau_{SA} = \pi/8$. 
As is evident, the engine charges the ancillas, leading to a final ensemble with clearly larger ergotropy. 

In Fig.~\ref{fig:example}(b) we show the average ergotropy as a function of $g\tau_{SA}$, where it is evident that stronger interactions lead to monotonic improvements in the charging process.
This is expected since higher $g\tau_{SA}$ imply more information is available to the demon to make the decision.
We also show, for comparison, the ergotropy which would be obtained if all ancillas were to be processed by the engine, irrespective of the measurement outcomes (labeled ``engine''). 
In this case the interaction with the system causes an overall degradation of $\mathcal{W}$. 
This happens because the interaction~\eqref{example_HSA} dephases the ancillas. 
Hence, the coherent part of the ergotropy tends to be lost (while the population part is unaffected). 

\begin{figure}
    \centering
    \includegraphics[width=0.45\textwidth]{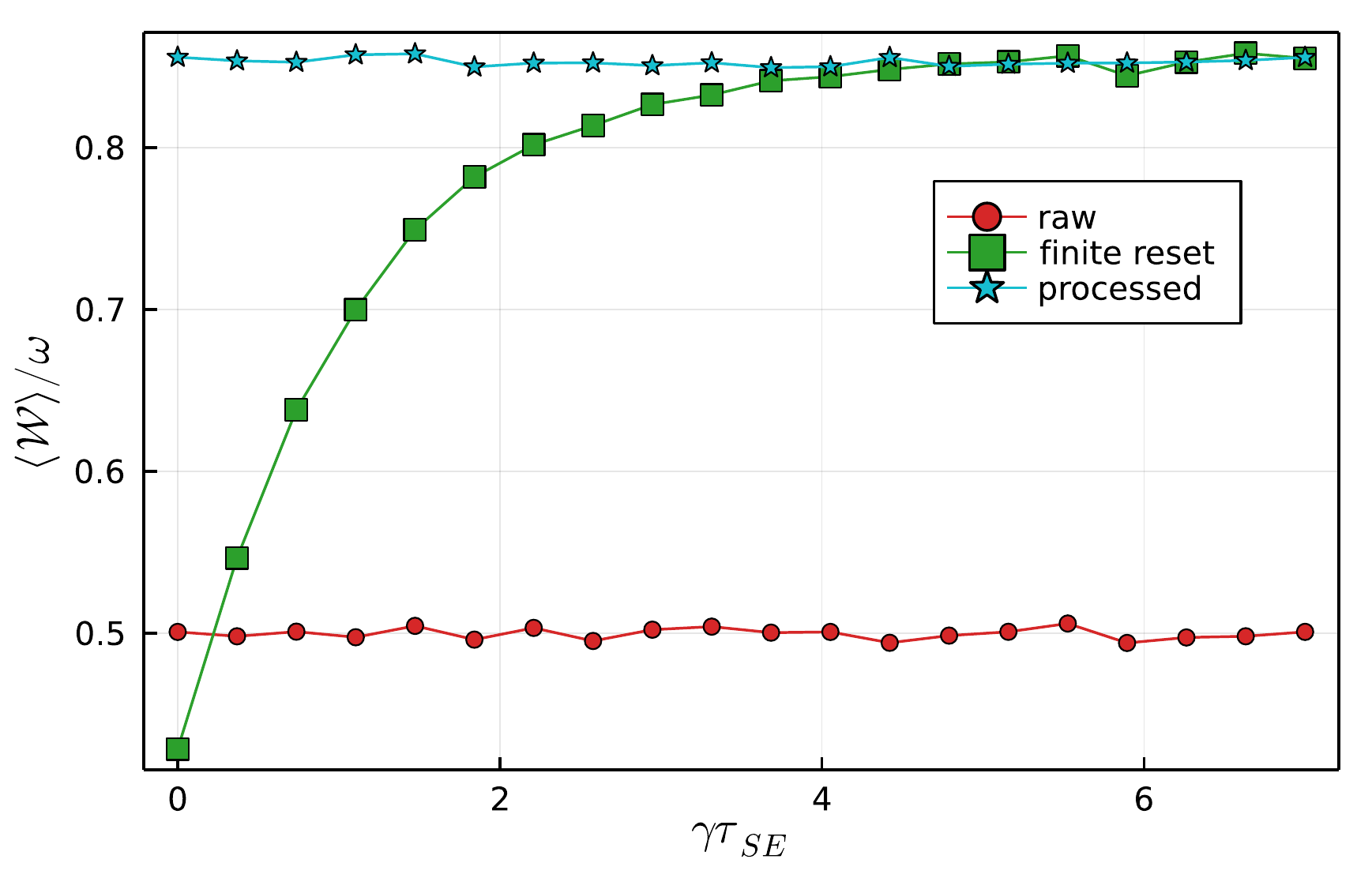}
    \caption{The curve marked ``finite reset'' depicts the dependence  of the average ergotropy on the system relaxation time $\gamma\tau_{SE}$. 
    The data was sampled from   $N = 10^4$ simulations, with the system-ancilla interaction strength  fixed at $g\tau_{SA} = \pi/8$.
    The other two curves, marked  ``raw'' and ``processed,''  are shown for comparison, and are similar to those from Fig.~\ref{fig:example}(b).
    }
    \label{fig:example2}
\end{figure}

Next we investigate what happens when the state of the ancilla is not fully reset after each step. 
Due to the projective nature of the measurement, after each collision the system will either be in $|+\rangle$ or in $|-\rangle$. 
The state, after a time $\gamma\tau_{SE}$, under the action of Eq.~\eqref{QME}, will thus be 
\begin{equation}
    \rho_{S|\pm}(t) = \begin{pmatrix}
    1 - e^{-\gamma\tau_{SE}}/2 & \pm e^{-\gamma\tau_{SE}/2 + i \omega_S \tau_{SE}}/2
    \\[0.2cm]
    \pm e^{-\gamma\tau_{SE}/2- i \omega_S \tau_{SE}}/2 & e^{-\gamma\tau_{SE}}/2
    \end{pmatrix},
\end{equation}
which are thus taken as the initial states of the next collision.
Results for the average ergotropy are shown in Fig.~\ref{fig:example2}. 
As can be seen, when $\gamma\tau_{SE}$ is finite, the ergotropy is gradually reduced. 
This happens because when the system is not properly erased, it affects the demon's ability to make proper decisions. 
In fact, if $\gamma\tau_{SE}$ is very small, one can even obtain an average ergotropy which is worse than that of a fully random ensemble. 

\section{\label{sec:energetics}Energetics}

We now discuss in further detail the energetics of the problem. 
We divide the problem in 3 steps: interaction, measurement and conditional unitary pulse.
For simplicity, we focus on full system resets ($\gamma\tau_{SE} \to \infty$).
The interaction~\eqref{example_HSA} does not affect the energy of the ancillas since $[H_{SA}, H_A] = 0$. 
But it does affect the energy of the system. 
The net change in energy of system plus ancilla, in one collision, assuming the ancilla is in $\psi_A$, is thus given by 
\begin{equation}
    \Delta E_{\rm col} = \tr\Big\{(\rho_{S|\psi_A} - \rho_S) H_S\Big\}.
\end{equation}
This change reflects the inherent work cost associated to the interaction $H_{SA}$, known as on/off work~\cite{DeChiara2018,Landi2021a}.
Notice, however, that this will depend on the Hamiltonian in the system, which has a generic gap $\omega_S$ (not necessarily resonant with the ancilla's gap $\omega$). 
The on/off work can thus be made arbitrarily small by choosing $\omega_S$ to be small. 
This  means that it is well possible to operate the engine in a regime where the energy cost of the collision is negligible. 

Next we turn to the effects of the measurement. 
We assume that the ancilla's initial state has the generic form 
$|\psi_A \rangle = \cos(\theta/2)|0\rangle + e^{i \phi} \sin(\theta/2)|1\rangle$. 
The average energy of the ancillas after the measurement, given outcomes $x = \pm1$, will then be 
\begin{equation}
    E_{A|x} = -\frac{\omega}{2} \frac{\cos\theta + x \sin(2g\tau_{SA})}{1 + x \cos\theta \sin(2g\tau_{SA})}.
\end{equation}
Averaging this over the probabilities~\eqref{example_likelihood} recovers the initial average energy $\langle \psi_A | H_A |\psi_A\rangle$.
Thus, up to this point, no work is performed in the ancillas (on average). 

The actual work comes from the controlled unitary pulse, which is applied only when $x = + 1$. 
This causes the energy of the ancillas to change to 
\begin{equation}
    \tilde{E}_{A|+1} = \tr\Big\{\sigma_x \rho_{A|+1} \sigma_x H_A\Big\} = -E_{A|+1}.
\end{equation}
The net work is therefore 
\begin{equation}
    \mathbb{W}_+ = \tilde{E}_{A|+1} - E_{A|+1} = \omega \frac{\cos\theta +  \sin(2g\tau_{SA})}{1 +  \cos\theta \sin(2g\tau_{SA})},
\end{equation}
which $\mathbb{W}_- = 0$ when $x = -1$.
The average work is thus
\begin{equation}
    \mathbb{W} = P(+1|\psi_A) \mathbb{W}_+ + P(-1|\psi_A) \mathbb{W}_- = \frac{\omega}{2} (\cos\theta + \sin(2g\tau_{SA})).
\end{equation}
Notice how work is still performed even if the system and ancilla do not interact ($g\tau_{SA} = 0$).
This happens because, even though they don't interact, we assume that the system is nonetheless still measured, thus yielding equally likely outcomes $x = \pm 1$.
That is to say, half of the time the pulse is applied. 

We now analyze this from the perspective of the ergotropy. 
The initial ergotropy is $\mathcal{W}_0 = \omega \sin^2(\theta/2)$. 
After the measurements (but before the pulse), the ergotropies conditioned on each outcome are 
\begin{equation}
    \mathcal{W}_x = \mathcal{W}(\rho_{A|x,\psi_A}) = \omega \sin^2(\theta/2) \frac{1-x \sin(2g\tau_{SA})}{1+ x \cos\theta \sin(2g\tau_{SA})}.
\end{equation}
Since the measurement does not perform any work, on average, we simply have $\sum_x P(x|\psi_A) \mathcal{W}_x = \mathcal{W}_0 = \omega \sin^2(\theta/2)$, as it must be.

When the pulse is performed, however, the ergotropy changes to 
\begin{equation}
    \tilde{\mathcal{W}}_+ = \omega \cos^2(\theta/2) \frac{1+ \sin(2g\tau_{SA})}{1+ \cos\theta \sin(2g\tau_{SA})}.
\end{equation}
The net change in ergotropy is, of course, the work injected, 
\begin{equation}
    \tilde{\mathcal{W}}_+ - \mathcal{W}_+ = \mathbb{W}_+.
\end{equation}
The final average  ergotropy is then 
\begin{IEEEeqnarray}{rCl}
\mathcal{W}_{\rm processed} &=& P(+1|\psi_A) \tilde{\mathcal{W}}_+ + P(-1|\psi_A) \mathcal{W}_- 
\nonumber\\[0.2cm]
&=& \frac{\omega}{2} \Big[ 1 + \sin(2g\tau_{SA})\Big].
\end{IEEEeqnarray}
% $\mathcal{W} = \sum\limits_x P(x|\psi) \mathcal{W}_x$, is then
% \begin{equation}
%     \mathcal{W} = \frac{\omega}{2} \Big[ 1 + \sin(2g\tau_{SA})\Big].
% \end{equation}
If $g\tau_{SA}$, this reduces to $\omega/2$, which is half the maximum value it may have. 
Thus, if the machine is applied under no information about the ancillas whatsoever, it would result in an average ergotropy of $\omega/2$. 
And if $g\tau_{SA} = \pi/$, the average ergotropy achieves its maximum value $\omega$. 
This therefore fully accounts for the behavior observed in  Fig.~\ref{fig:example}.

\section{Discussion}

In this paper we put forth the idea of an autonomous engine, which processes random incoming ancillas with the goal of increasing their ergotropy.
There are endless possible variations of such an engine that one might construct.
The goal of the present proposal was to build a minimal engine, where the basic effects could be made evident. 
In particular, they are the following. 
First, the idea that, in reality, ancillas are usually sampled from an ensemble of pure states. 
Collision models often assume that the ancillas arrive in mixed states $\rho_A$, which could be viewed as the ensemble average. 
But for the present purposes, it is much more realistic to assume that in each collision, the state of the ancilla is pure, but not necessarily known.
In fact, for the example in Fig.~\ref{fig:example} the ensemble average would be simply the identity $\rho_A = I_A/2$. 

The second relevant aspect of this construction is the need for the state of the system to be properly reset after each step, as it plays the role of a memory. 
If this is not done, the ability of the demon in making a decision based on the measurement outcomes is severely degraded, as Fig.~\ref{fig:example2} illustrates very clearly. 

Finally, the third relevant point is the energetic balance of the problem. 
This has long been a major advantage of collisional models, as it enables for precise accounting of all possible energy sources and sinks. 
The analysis in Sec.~\ref{sec:energetics} showed how this can be used to pinpoint, at the level of each possible measurement outcome, whether or not work is being performed, and how this affects the ergotropy at each step.

\section*{Acknowledgments}

The author acknowledges the financial support of the S\~ao Paulo Funding Agency FAPESP (Grant No.~2019/14072-0.), and the Brazilian funding agency CNPq (Grant No. INCT-IQ 246569/2014-0).

\bibliography{library}

\end{document}